\documentclass[conference]{IEEEtran}
\IEEEoverridecommandlockouts
\usepackage{amsmath,amssymb,amsfonts}
\usepackage{algorithmic}
\usepackage{graphicx}
\usepackage{textcomp}
\usepackage{xcolor}
\usepackage{placeins}

\usepackage{hyperref}

\usepackage{pdfpages}

\def\BibTeX{{\rm B\kern-.05em{\sc i\kern-.025em b}\kern-.08em
    T\kern-.1667em\lower.7ex\hbox{E}\kern-.125emX}}

\begin{document}

\title{Performance Analysis of an Optimization Algorithm for Metamaterial Design on the Integrated High-Performance Computing and Quantum Systems}

\author{
\IEEEauthorblockN{Seongmin Kim}
\IEEEauthorblockA{National Center for Computational Sciences \\
Oak Ridge National Laboratory\\
Oak Ridge, Tennessee, USA \\
kims@ornl.gov}
\and
\IEEEauthorblockN{In-Saeng Suh}
\IEEEauthorblockA{National Center for Computational Sciences \\
Oak Ridge National Laboratory\\
Oak Ridge, Tennessee, USA \\
suhi@ornl.gov}
}

  

\maketitle

\begin{abstract} 
Optimizing metamaterials with complex geometries is a big challenge. Although an active learning algorithm, combining machine learning (ML), quantum computing, and optical simulation, has emerged as an efficient optimization tool, it still faces difficulties in optimizing complex structures that have potentially high performance. In this work, we comprehensively analyze the performance of an optimization algorithm for metamaterial design on the integrated HPC and quantum systems. We demonstrate significant time advantages through message-passing interface (MPI) parallelization on the high-performance computing (HPC) system showing approximately 54\% faster ML tasks and 67 times faster optical simulation against serial workloads. Furthermore, we analyze the performance of a quantum algorithm designed for optimization, which runs with various quantum simulators on a local computer or HPC-quantum system. Results showcase $\sim$24 times speedup when executing the optimization algorithm on the HPC-quantum hybrid system. This study paves a way to optimize complex metamaterials using the integrated HPC-quantum system.
\end{abstract}


\section{\textbf{Introduction}}
Metamaterials derive their properties from the inherent properties of constituent materials as well as the geometrical arrangements of sub-wavelength-scaled meta-atoms \cite{meinzer2014plasmonic}, enabling them to manipulate incident waves \cite{cummer2016controlling}. Over the past decade, metamaterials have drawn considerable interest as potential substitutes for conventional optical, mechanical, or thermal materials due to their unique properties, expecting them increasingly applicable in practical settings with decreasing optimization and fabrication costs \cite{kim2023scalable, wang2020thermal, liu2017concepts}. While optimizing their geometrical features is important to achieve high performance for practical applications, their complex parametric spaces lead to an explosion of optimization spaces, making optimization processes impractical in realistic time scales \cite{kitai2020designing, kim2023design}. Furthermore, evaluating their properties often requires considerable time, even with approximation or numerical methods, resulting in significant computational costs. These difficulties make the optimization of metamaterials challenging works. 

An {\it active learning algorithm} \cite{kitai2020designing,kim2023design,doi:10.1021/acsenergylett.2c01969,alexeev2023quantumcentric}, combining machine learning (ML), quantum computing, and wave-optics simulation in an iteration, has been proven to be highly efficient in designing various kinds of photonic structures including metamaterials \cite{kitai2020designing, doi:10.1021/acsenergylett.2c01969, kim2024wide}. This algorithm keeps enhancing the quality of datasets throughout the optimization process by iteratively adding higher-quality data points, leveraging benefits of ML- and quantum computing-assisted optimization processes. As a result, it can build more accurate surrogate models based on high-quality datasets, enabling the identification of the global optimum or near-optimum metamaterial structure. The success of the active learning algorithm has been demonstrated across various applications, such as metamaterial radiator \cite{kitai2020designing}, metamaterial thermal emitter \cite{wilson2021machine}, metamaterial solar absorber \cite{kim2023design}, metamaterial optical diode \cite{kim2024quantum}, transparent radiative cooler \cite{doi:10.1021/acsenergylett.2c01969}, and wide-angle spectral filter \cite{kim2024wide}. 

However, previous works have largely focused on simple geometries with relatively small parametric spaces due to computational limitations on conventional computing. With growing interest in designing more complex materials with potentially higher performance, there is a need to explore advanced computational techniques. Leveraging high-performance computing (HPC) for parallel computations presents an efficient solution to address these challenges. By distributing workloads across multi-processors, HPC can accelerate overall computing processes and handle large-scale problems that are intractable on local computers \cite{unat2017trends, haghi2021workload}. Therefore, it is important to analyze the performance of parallel computations on an HPC system to prepare an HPC-enhanced optimization algorithm for complex metamaterial design.

In the meantime, the active learning algorithm involves quantum computing to solve a surrogate model represented by a Hamiltonian formulation (quadratic unconstrained binary optimization; QUBO) \cite{kitai2020designing, kim2023design, doi:10.1021/acsenergylett.2c01969, kim2024wide}. While previous studies mostly relied on quantum annealing, a specialized quantum computing technique for solving combinatorial optimization problems, the utilization of gate-based quantum computers has become more important due to their universal applicability across various problem domains \cite{fellner2022universal, fankhauser2023multiple, biamonte2021universal}. In this regard, the quantum approximate optimization algorithm (QAOA) is considered a valuable quantum algorithm for universal gate-based quantum computers in the noise intermediate-scale quantum (NISQ) era \cite{bechtold2023investigating}. 

QAOA can efficiently solve combinatorial optimization problems by iteratively employing classical and quantum computing, resulting in a global optimum or near-optimum solution \cite{ushijima2021multilevel, khairy2020learning}. However, performance evaluation of this quantum algorithm (QAOA) on quantum hardware is hard due to limited resources \cite{doi2020cache}; thus, quantum simulators have become essential tools for executing and evaluating quantum algorithms \cite{daley2022practical, altman2021quantum}. Quantum simulators have varying performances depending on their types due to their inherently different operating mechanisms, thereby leading to different outputs on each simulator even for the same quantum algorithm \cite{chundury2023peps, suh2024simulating}. In addition, an HPC-quantum hybrid system, such as the Oak Ridge Leadership Computing Facility (OLCF) Frontier supercomputer which is loosely integrated with the Quantum Computing User Program (QCUP) quantum resources \cite{qcup}, can offer enhanced capabilities for quantum simulators to solve optimization problems with QAOA. Therefore, it is required to evaluate the performance of QAOA across various quantum simulators on a local computer or HPC-Quantum system to understand which simulator is the most efficient depending on a given problem.

In this work, we comprehensively analyze the performance of an optimization algorithm aimed at designing highly complex metamaterials on the integrated HPC and quantum systems. This work makes the following contributions:
\begin{itemize}
\item  We provide clear insights into under what conditions the optimization processes utilizing the HPC and quantum system have advantages over those utilizing conventional computing.
\item  We demonstrate the significant benefits of employing parallel computing using the message-passing interface (MPI) on multi-processors for wave-optics simulations.
\item  We showcase the acceleration in ML training using MPI on multi-processors, and the potential advantages by employing graphic processing units (GPUs) for ML tasks.
\item  We verify that the performance of QAOA varies across different quantum simulators, highlighting the potential advantages of utilizing the HPC-quantum hybrid system to solve combinatorial optimization problems with QAOA.
\end{itemize}

The remainder of this paper is organized as follows: Section 2 briefly introduces the active learning algorithm for metamaterial design. Section 3 explains the current limitations in each step of the active learning algorithm and presents solutions to overcome those challenges. Section 4 shows detailed performance analysis results including our discussion. We close this paper with the conclusion in Section 5.

\section{\textbf{Brief Background of an Active Learning Algorithm for Metamaterial Design}}
The active learning algorithm \cite{kitai2020designing,kim2023design,doi:10.1021/acsenergylett.2c01969,alexeev2023quantumcentric} for metamaterial design involves three key components:
\begin{enumerate}
\item  {\it Machine Learning:} ML is employed to construct a surrogate model by learning the relationship between inputs and corresponding output performance metrics (i.e., figure-of-merit; FOM). This surrogate model serves as a metamodel representing the relationship between metamaterial structures and their FOMs.
\item {\it Quantum Computing:} Quantum computing is employed to solve a given surrogate model generated by the ML model. Here, variational quantum algorithms, such as QAOA, can be used to efficiently solve given optimization problems leveraging gate-based quantum computing resources \cite{ushijima2021multilevel}. By formulating the problem as a quantum optimization task, quantum computing can efficiently explore the solution space and identify promising metamaterial structures. 
\item {\it Simulation:} Wave-optics simulation is used to evaluate the performance of the metamaterial structures identified by the ML-quantum computing loop. The simulation calculates FOM associated with the identified metamaterial structure, providing feedback for the optimization process.
\end{enumerate}

After each iteration, a new data point is added to the dataset that consists of binary vectors representing metamaterial structures identified by quantum computing and their corresponding FOMs calculated from wave-optics simulation \cite{kim2024wide, doi:10.1021/acsenergylett.2c01969}. The iterations enable the inclusion of high-quality data points, which in turn allows ML to refine the surrogate model. As a result, quantum computing can identify a more valuable metamaterial structure based on the improved surrogate model, leading to the identification of global or near-global optimal structures \cite{kim2024wide, doi:10.1021/acsenergylett.2c01969, kitai2020designing, kim2023design}.

\begin{figure}[h]
\centering
\includegraphics[width=1.0\linewidth]{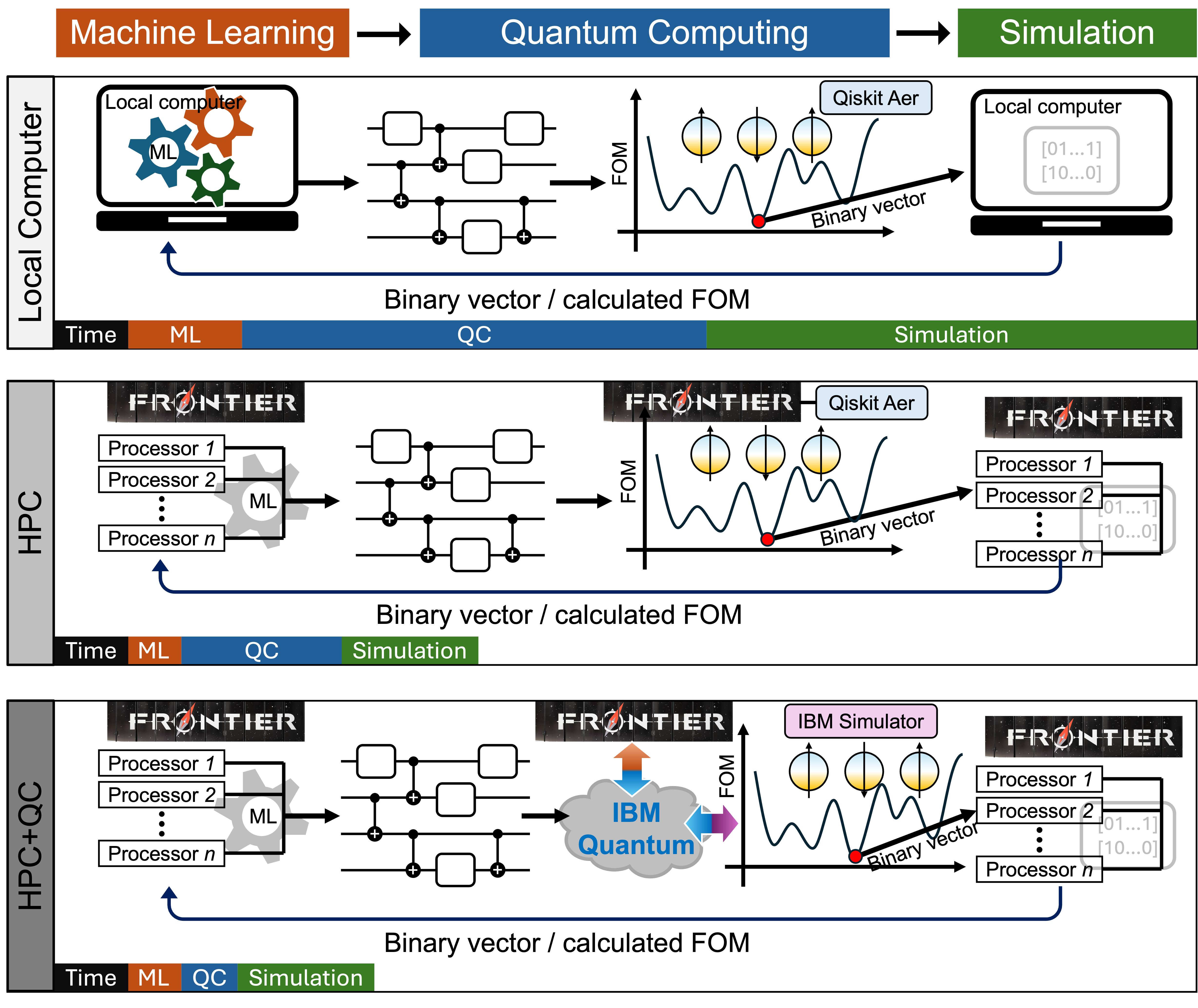}
\caption{Schematic of the active learning algorithms on different computing systems, showing the key components (i.e., ML, quantum computing (QC), simulation) and workflow. Bars at the bottom of each computing system represent the approximate required time for each component. The HPC (middle) and integrated system (HPC+QC, bottom) show significant acceleration in computational time compared to a local computer (top).}
\label{fig:fig-1}
\end{figure}

Despite the demonstrated efficiency of the active learning algorithm, computational limitations exist for each component with conventional computing, especially when designing complex metamaterials. These challenges can be overcome by leveraging advanced computational capabilities, such as HPC and HPC-quantum integrated systems (Figure \ref{fig:fig-1}). In the following section, we define complex optimization problems and introduce strategies to overcome the computational limitations, paving the way for more efficient and scalable optimization processes.

\section{\textbf{{Strategies to Solve Large-Scale Optimization Problems}}}
\subsection{Optimization Problem - Radiative Cooling Materials}
Metamaterials are applicable to a wide range of fields including optical or thermal materials such as radiative cooling materials \cite{kim2024wide, doi:10.1021/acsenergylett.2c01969}, optical diodes \cite{kim2024quantum}, and thermophotovoltaics \cite{kim2023design}; One of the recent popular applications is radiative cooling techniques \cite{doi:10.1021/acsenergylett.2c01969}. The radiative cooling technique enables the emission of thermal radiation through the atmospheric window (AW; wavelength: 8 to 13 um) to outer cold space ($\sim$ 3 K) without consuming energy and refrigerants \cite{li2019radiative}. Given the growing efforts to address global warming in the last few decades, this technique has drawn great attention as a green and passive cooling solution. In particular, transparent radiative coolers (TRCs) can be applied to windows which largely contributes to energy loss in enclosed spaces such as buildings \cite{wang2021scalable}. Photonic structures, such as optical metamaterials and multilayered structures, have been developed for TRC applications \cite{kim2021visibly, dang2022ultrathin}. These materials exhibit high transmissivity in the solar spectrum range and high emissivity in the AW. However, ultraviolet (UV) and near-infrared (NIR) light from sunlight significantly contribute to optical heating, thus they are desired to be reflected to maximize TRC performance \cite{kim2024wide, doi:10.1021/acsenergylett.2c01969}. 

Many efforts have been made to develop high-performance TRCs using various optimization algorithms including the active learning algorithm. However, computational limitations have often confined TRC designs to relatively simple geometries, hindering the exploration of more complex designs that could offer better performance. For example, although 1,000-layered structures can have high TRC performance, the optimization processes involve high computational costs, hindering the optimization with such complexity in realistic time scales. Furthermore, computational costs significantly grow with an increasing number of simulation conditions; for instance, considering TRC performance across a wide range of incident angles can further increase computational costs. In this scenario, parallel computing on HPC systems can be greatly useful to mitigate these challenges.

\subsection{Machine Learning}
The active learning algorithm integrates iterative processes of ML and quantum computing to accelerate optimization tasks. In this algorithm, an ML constructs a surrogate model, which serves as a representation of the optimization problem. Quantum computing then operates on this surrogate model formulated with a Hamiltonian (e.g., quadratic unconstrained binary optimization; QUBO). This process aims to identify an optimal binary vector, which corresponds to the ground state of a given surrogate model represented by the QUBO formulation. One of the keys of this algorithm is the factorization machine (FM), a supervised machine learning model. FM enables to describe the relationship between input binary vectors ($\textbf{\textit{x}}$) and outputs (y) through linear and quadratic coefficients, which is given by the following equation \cite{rendle2010factorization}:
\begin{equation}
y = w_0 + \sum_{i=1}^{n} w_i {\textit{x}}_i + \frac{1}{2} \sum_{f=1}^k \left[
\left( \sum_{i=1}^n v_{i,f} {\textit{x}}_{i} \right)^2 - \sum_{i=1}^{n} v_{i,f}^2 {\textit{x}}_{i}^2 \right]
\label{eqn4}
\end{equation}
where ${\textit{x}}_{i}$ represents an $i$-th element (either 0 or 1) of $\textbf{\textit{x}}$ with a length of $n$. The parameters $w_0, w_i, v_{i,f}$ and $k$ represent a global bias, linear coefficient, quadratic coefficient and latent space size, respectively. One advantage of using FM in this algorithm is the transparency of this ML model. Hence, the model parameters can be employed to build a surrogate model for inverse design and combined with other methods, such as quantum computing, for global optimization within the active learning algorithm \cite{kitai2020designing, kim2024wide, doi:10.1021/acsenergylett.2c01969}. In addition, FM can computationally efficiently learn pairwise interactions by factorizing them, resulting in decreased algorithmic complexity for training from $O(k n^2)$ to $O(k n)$ \cite{rendle2010factorization}, as seen in the above equation. However, the optimization of complex metamaterial usually requires lots of data points, and training with a large volume of data can be computationally costly, requiring significant time to complete. 

A mini-batch is a subset of the entire dataset commonly used in ML training. Using mini-batches provides several advantages, including computational efficiency, smoother convergence, better exploration of parameter space, and mitigation of over/underfitting \cite{smith2019super}. Furthermore, mini-batch training allows for parallelization on GPU or HPC clusters, potentially accelerating the training process \cite{lee2019improving}. However, it is important to consider communication overheads in parallel computing \cite{sattar2019overcoming, marjanovic2010overlapping}, which refer to the additional time required for coordinating and exchanging data between parallel computing units, such as cores or nodes in a cluster. Communication overheads can sometimes outweigh the advantages of parallel computing. To investigate the advantage of utilizing parallel computing, we systematically compare FM training time using different volumes of datasets on different computing systems; The training experiments are conducted on a single-processor, GPU, and multi-processors.

We generate datasets to train FM with sizes of $n_{d}$ $\times$ $n_{t}$ for this study, where $n_{d}$ is the number of decision variables and $n_{t}$ represents the total number of data points. For example, a dataset with a size of 120 $\times$ 1,000 indicates that it has 120 decision variables (i.e., binary bit-string length of 120) and 1,000 data pairs (consisting of 1,000 binary vectors and 1,000 corresponding calculated FOMs). An Intel® Xeon® Gold 6248R CPU is used to train FM on a single processor. For parallel computing, we use the OLCF Frontier for MPI-based parallel processing, and the OLCF Jupyter GPU Lab, which offers a setup with 16GB Memory and Nvidia V100 GPU, for GPU-accelerated computations. We utilize the HPC system (OLCF Frontier), which features 64-core AMD “Optimized 3rd Gen EPYC” CPUs, to implement MPI for parallel computations. The tests are conducted on a single node unless otherwise noted. 

\subsection{Quantum Approximate Optimization Algorithm}
As stated above, FM model parameters are well-fitted to the QUBO formulation, represented by $n \times n$ upper-triangular matrix $\mathbf{Q}$, where the objective function can be defined as the following:
\begin{equation}
 \bar{y} = \sum_{\textbf{\textit{x}}_{i} \in \{0,1\}^n} \textbf{\textit{x}}^T \mathbf{Q} \textbf{\textit{x}}
\label{eqn5}
\end{equation}
The objective of quantum computing is to find an optimal binary vector $\bar{\textbf{\textit{x}}}$ that minimizes the expected output ($\bar{y}$):
\begin{equation}
\bar{\textbf{\textit{x}}} = {\rm arg ~ \min_{\textit{x}}} ~ \bar{y}
\label{eqn6}
\end{equation}
Combinatorial optimization problems, such as max-cut, traveling salesman, graph partitioning, and metamaterial optimization problems, are known to be NP-hard, posing explosive searching spaces with an increasing number of decision variables for optimization \cite{li2018combinatorial, bittel2021training}. Therefore, these problems become intractable in classical computing systems when many decision variables are considered. Quantum computing that can evaluate numerous possible combinations simultaneously is highly promising for such combinatorial optimization problems. 

QAOA is a variational quantum algorithm specially designed to solve combinatorial optimization problems on universal gate-based quantum computers, leveraging the limited quantum computing resources (e.g., a limited number of qubits, and errors from noise and imperfect gates) in the NISQ era \cite{bittel2021training, bechtold2023investigating}. This quantum algorithm can converge to high-quality solutions close to the global optimum by iteratively employing both classical and quantum computing \cite{ushijima2021multilevel, khairy2020learning}. Although QAOA may not outperform classical optimization methods for all problems due to the limitation of the current quantum hardware, it is considered highly valuable as a practical tool for early quantum machines to understand the capability of quantum computing \cite{hastings2019classical, bechtold2023investigating}.

While quantum computers have demonstrated potential advantages over classical computers in several fields, a lack of accessibility to real quantum devices poses considerable difficulties in executing quantum algorithms \cite{daley2022practical, altman2021quantum}. In this regard, quantum simulators play a significant role in simulating and analyzing quantum mechanisms to solve given quantum problems or quantum circuits. In this work, we choose to use quantum simulators from IBM Quantum (e.g., ibmq$_{-}$qasm$_{-}$simulator, simulator$_{-}$statevector, simulator$_{-}$mps) and AerSimulator on a local computer. These simulators can be used to run quantum algorithms on a local computer or HPC system through Qiskit, or by submitting jobs on a local or HPC-quantum hybrid system \cite{ravi2021adaptive}. 

The operational mechanisms of each quantum simulator are different, resulting in different outputs even for the same quantum algorithm \cite{chundury2023peps, suh2024simulating}. In addition, quantum simulators may perform better when jobs are submitted or executed on the integrated HPC and quantum systems. Hence, we comprehensively analyze the performance of each simulator for a specific quantum algorithm (i.e., QAOA) executed or submitted on a local computer (Apple M2 Max, 32 GB Memory) or HPC-quantum system (OLCF Frontier). We establish various QUBO matrices with different problem sizes (i.e., numbers of decision variables ($n$) from 4 to 32; $n$ $\times$ $n$ upper-triangular matrices fully filled with random real numbers). We solve these QUBOs using QAOA with different simulators on the local computer and HPC-quantum system. We then analyze the accuracy and time-to-solution of each method. The accuracy is calculated by dividing the obtained solution by the known global optimum value (Accuracy = solution / global optimum). The time-to-solution is measured from the start to the end of QAOA jobs, including factors such as queue time and communication between the local computer and IBM server \cite{ravi2021quantum}.

\subsection{Transfer Matrix Method}
Transfer matrix method (TMM) is a promising mathematical technique to analyze the behavior of waves when they propagate through layered structures \cite{kim2024wide, doi:10.1021/acsenergylett.2c01969}. TMM needs to define the overall transfer matrix ($\mathbf{M}$) used to calculate the transmission and reflection coefficients for the system, which is composed of propagation matrices ($\mathbf{P}$) and transfer matrices ($\mathbf{T}$). $\mathbf{P}$ is defined based on the properties of a layer and wavelength of the incident wave, describing how the wave behaves as it propagates through a layer, which is given by \cite{katsidis2002general}:
\begin{equation}
    \mathbf{P} =
 \left[ \begin{array}{cc}
 e^{ik_{w}d} & 0 \\
 0        & e^{-ik_{w}d} 
 \end{array}
 \right]
 \label{eqn1}
\end{equation}
where ${k_{w}}$ is a wavevector in the layer ($k_{w} = \frac{2 \pi n_{r}}{\lambda}$), and $n_{r}$ and $d$ represent the refractive index and thickness of a layer. 

The transfer matrix $\mathbf{T}$ describes how the wave propagates and interacts when it passes from one layer to another, explaining amplitudes and their derivatives at the interfaces between adjacent layers (e.g., layer $j$ to layer $j+1$ refers to ${\rm T_{j,j+1}}$), as the following \cite{katsidis2002general}:
\begin{equation}
    \mathbf{T}_{j,j+1} =
 \left[ \begin{array}{cc}
 t_{j,j}     & t_{j,j+1} \\
 t_{j+1,j}   & t_{j+1,j+1} 
 \end{array}
 \right]
 \label{eqn2}
\end{equation}
Here, $t_{j,j}$ and $t_{j,j+1}$ respectively represent the transmission amplitude for the wave component parallel to the interface and perpendicular to the interface. $t_{j+1,j}$ and $t_{j+1,j+1}$ respectively represent the reflection amplitude for the wave component parallel to the interface and perpendicular to the interface. They are calculated based on incident angles, polarization states, and optical properties of materials in layers.

$M$ is obtained by cascading $P$ and $T$ for each layer, given by \cite{katsidis2002general, li2007second}:
\begin{equation}
    \mathbf{M} = \mathbf{T}_{N-1,N} \times \mathbf{P}_{N} \times \ldots \times \mathbf{T}_{1,2} \times \mathbf{P}_{1} 
\label{eqn3}
\end{equation}
TMM is particularly useful for calculating the optical properties (transmission, reflection, and emission) of multilayered systems. However, it faces challenges when evaluating the properties of highly complex structures under many simulation conditions, particularly due to the need to calculate transfer matrices serially for each wavelength domain and each simulation condition. This results in a significant computational burden, especially when considering many simulation conditions (e.g., wide wavelength regime and wide incident angle). 

\begin{figure}[!ht]
\centering
\includegraphics[width=1.0\linewidth]{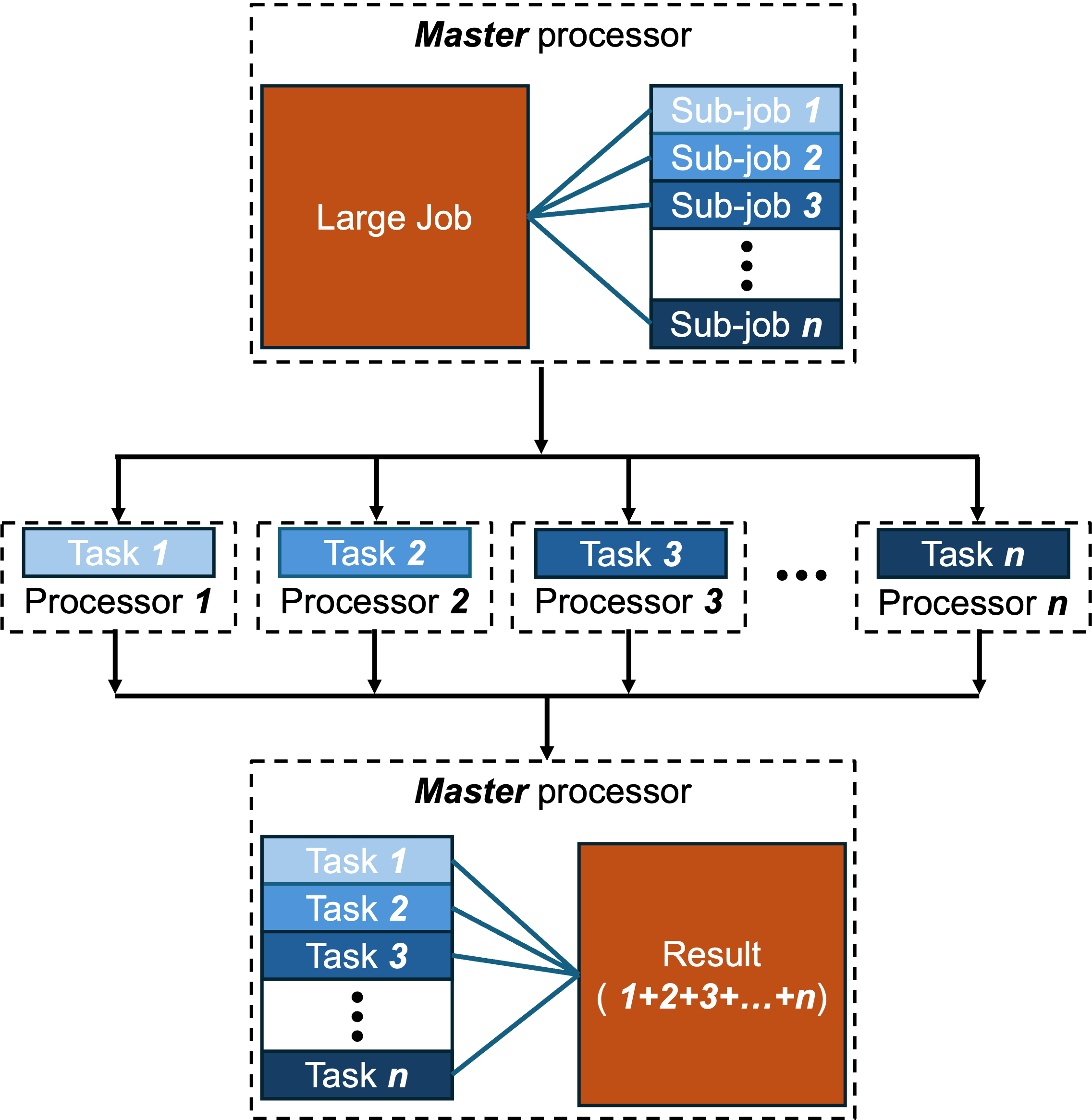}
\caption{Schematic of MPI parallelization of TMM on HPC systems.}
\label{fig:fig-2}
\end{figure}

To address these challenges, we implement MPI to parallelize TMM simulations on HPC systems. By using multi-processors, MPI allows us to efficiently distribute the computational workload and calculate the optical properties with TMM across wide wavelength domains and multiple simulation conditions. We decompose the wavelength domain into several sub-domains based on the number of MPI processors. Each processor takes the decomposed wavelength domain for the simulation, thereby reducing the workload. Hence, a large job (the whole wavelength domain) is split into sub-jobs (sub-wavelength domains), and each processor works with the sub-jobs. Then, the master processor calculates the optical property of the given optical system by taking all results from each processor (Figure \ref{fig:fig-2}).

In this work, we evaluate the performance of HPC-enhanced TMM implementing MPI by taking an example of a 1,000-layered photonic structure with a different number of simulation conditions (50 to 1,000, the incident angle from 0$^{\circ}$ to 89$^{\circ}$). Note that it is impractical to evaluate the optical characteristics of such a complex structure with 400 simulation conditions on a local computer (Apple M2 Max, 32 GB Memory) due to computational limitations. However, by leveraging HPC to utilize MPI parallelization for TMM, we can efficiently solve this highly complex problem within a minute, demonstrating a significant time advantage over the conventional counterpart. We utilize the OLCF Frontier for leveraging MPI parallel computations. The tests are conducted on a single node unless otherwise noted. 

\section{\textbf{Results}}
\subsection{Performance Analysis of HPC-Enhanced FM}
FM training is an essential step in the active learning algorithm, but it can pose computational challenges, particularly when optimizing complex metamaterial structures. Although FM training is known for its computational efficiency because of the reduced algorithmic complexity \cite{rendle2010factorization}, the training time can still be considerable especially when dealing with large volumes of data associated with complex metamaterial design. In this subsection, we investigate the performance of the FM training (i.e., FM training time) on different computing systems. 

First, we conduct a benchmarking study to verify the advantages of leveraging GPUs for parallel computations. We establish matrices of varying sizes, and measure the time required for matrix multiplications on different computing systems (CPU with a single processor and GPU for parallel computations). Figure 3A shows that the matrix operation on the CPU is faster than that on the GPU for small matrices (size less than 100 $\times$ 100). However, the GPU becomes increasingly efficient as the matrix size gets larger, and it is $\sim$15 times faster than the CPU for the calculation with a matrix size of 10,000 $\times$ 10,000 (takes 0.61 s on the GPU, but 9.94 s on the CPU). This benchmarking study clearly demonstrates that the GPU has a superior capability for matrix operations by taking advantage of parallel data processing, showing the potential advantages of GPU-accelerated ML training. Hence, we compare the times required for FM training on the CPU and GPU using different training datasets. 

As seen in Figure 3B, the CPU consistently outperforms the GPU for FM training tasks, primarily because the training datasets are not sufficiently large to fully exploit the advantages of parallel processing. Instead, the communication overhead associated with parallel computations outweighs the computational advantages. However, we observe that the discrepancy in the training time on the CPU and GPU decreases as the volume of training data increases. For example, the CPU is 17\% faster than the GPU (CPU: 11.98 s, GPU: 14.05 s) for training with a small dataset of 120 $\times$ 1,000, but this difference decreases to only 0.3\% for a larger dataset of 240 $\times$ 10,000 (CPU: 85.89 s, GPU: 86.17 s). These results illustrate that leveraging GPU-accelerated training may draw computational advantages by overcoming communication overhead when handling a large volume of data. Thus, it is expected that the training on the GPU will be faster than that on the CPU for datasets larger than 240 $\times$ 10,000, highlighting the potential for GPU acceleration in FM training for complex metamaterial design \cite{liang2018gpu}.

\begin{figure}[!ht]
\centering
\includegraphics[width=1.0\linewidth]{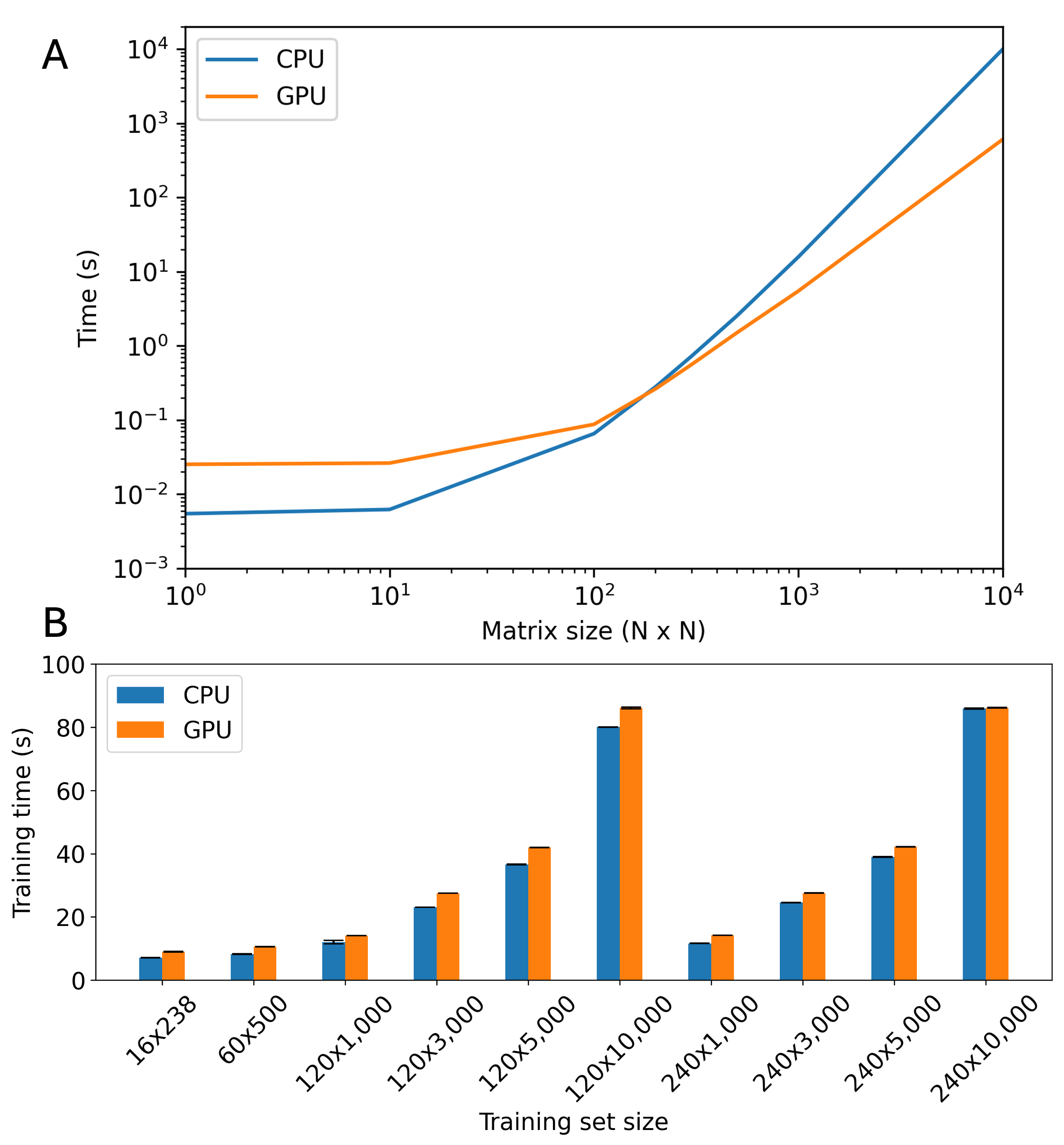}
\caption{\label{fig:fig-3}(A) Elapsed time to complete matrix multiplication conducted on the CPU and GPU as a function of matrix sizes. (B) FM training time with different volumes of datasets on the CPU and GPU.}
\end{figure}

Since FM has a relatively small number of model parameters and it can be trained on a relatively small dataset compared to other ML algorithms, it is generally difficult to observe acceleration in FM training on GPUs (Figure 3B). However, training time can be reduced through MPI parallelization on multi-processors instead of relying on a single processor \cite{xu2020distributed} (Figure 2). Hence, we investigate the performance of MPI-parallelized FM. Figure 4A presents the FM training time for different volumes of datasets with varying numbers of processors used for MPI. It shows some benefits of employing multi-processors on the HPC system (OLCF Frontier) when a dataset is small (e.g., 120 $\times$ 1,000). Hence, FM training takes 11.66 s, 8.10 s, 10.12, and 23.24 s when using CPU with single, 3, 6, and 21 processors, respectively. The training time on the single processor CPU grows more rapidly than that on the multi-processors, thus the advantage of MPI parallelization becomes more evident when a dataset is large. For instance, the training with a dataset size of 120 $\times$ 10,000 takes 79.76 s, 53.68 s, 53.93 s, and 67.09 s when using the CPU with single, 6, 9, and 21 processors, respectively. Note that it cannot be trained on the 3 processors, which may be due to the time required for MPI job allocation on the HPC system exceeding the wall time. Furthermore, we can infer that the communication overhead can outweigh the advantage achieved by utilizing multi-processors. Hence, for a dataset of 120 $\times$ 1,000, the most efficient training is achieved by using the 3 processors instead of using more processors (e.g., 6, 9, or 21 processors). Interestingly, using more processors is preferable when the volume of datasets increases, thus the training on the 6 processors is the most efficient for a dataset size of 120 $\times$ 10,000. The benefit of leveraging multi-processors is more obvious when datasets have larger volumes with more decision variables (e.g., 120 $\rightarrow$ 240); training with a dataset size of 240 $\times$ 10,000 takes 92.22 s, 59.98 s, 59.79 s, and 66.77 s when using the CPU with single, 6, 9, and 21 processors, respectively, demonstrating a 54\% acceleration by employing 9 processors compared to the training on the single-processor (Figure 4B). 

\begin{figure}[!ht]
\centering
\includegraphics[width=1.0\linewidth]{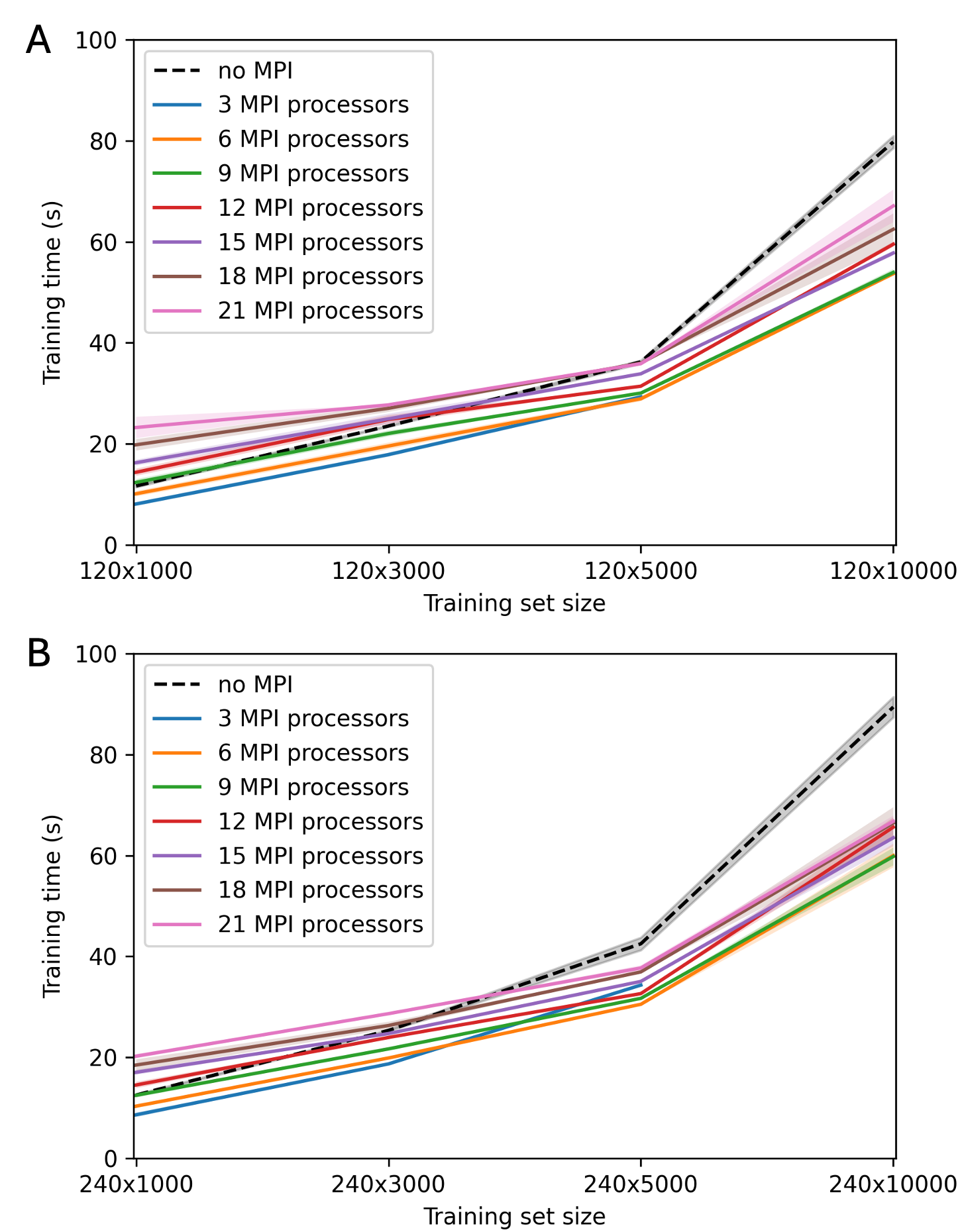}
\caption{\label{fig:fig-4}{FM training time on the Frontier with different numbers of processors used for MPI parallelization. Datasets include bit-string lengths of (A) 120 and (B) 240.}}
\end{figure}

Increasing the number of nodes while keeping the total number of MPI processors does not bring significant improvement in FM training time. We analyze the performance of HPC-enhanced FM on multi-nodes (keeping the total number of MPI processors to 21) with a training set size of 240 $\times$ 10,000 (Figure 5). The results show leveraging 4 nodes is beneficial with $\sim$10\% acceleration, but it does not bring further notable acceleration when using more than 4 nodes.

\begin{figure}[!ht]
\centering
\includegraphics[width=1.0\linewidth]{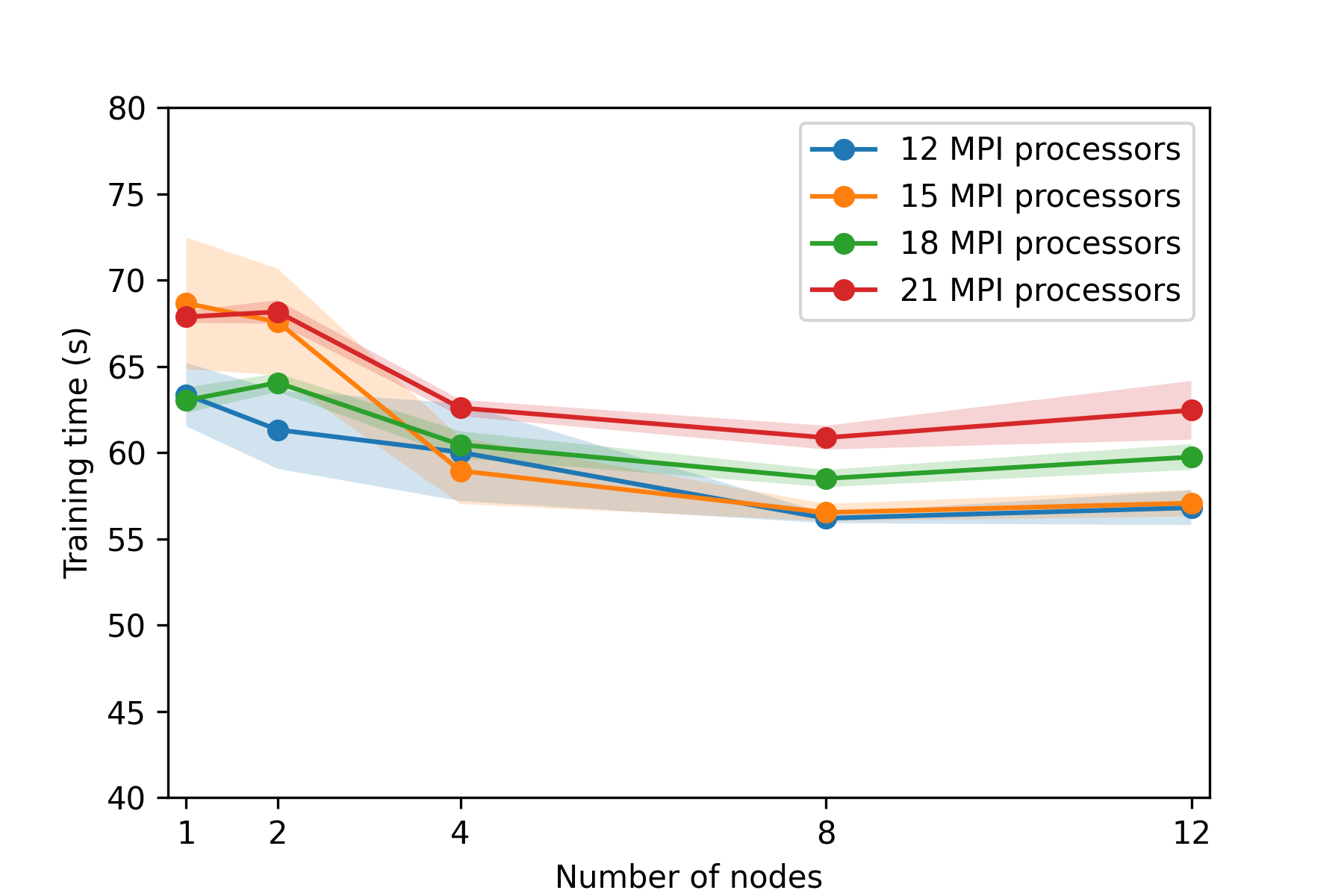}
\caption{\label{fig:fig-5}{FM training time on the Frontier with different numbers of nodes using MPI with multi-processors for a dataset size of 240 $\times$ 10,000.}}
\end{figure}

These results clearly verify the advantage of utilizing HPC-enhanced FM for MPI parallelization with multi-processors, especially when the volume of a dataset is sufficiently large. This advantage will be more significant with larger data volumes, allowing for faster training and efficient optimization of complex metamaterial structures that require large datasets for FM. Therefore, the HPC-enhanced FM will be integrated into the active learning algorithm for such optimization tasks.

\subsection{Performance Analysis of QAOA with Quantum Simulators}
QAOA is a promising quantum algorithm to solve combinatorial optimization problems with gate-based quantum computers, where an objective function is formulated with QUBO derived from FM model parameters \cite{kitai2020designing, kim2023design, doi:10.1021/acsenergylett.2c01969}. Here, as stated in the previous section, the performance of QAOA can vary depending on the type of quantum simulator used, thus we analyze the performance of QAOA on those simulators. IBM Quantum offers quantum simulators accessible via Qiskit, which can be run on a local computer (i.e., local simulators). These simulators include BasicAer simulator, and Aer simulator with different methods (Aer/density matrix, statevector, matrix product state (MPS), and automatic) \cite{doi2020cache}. Additionally, IBM Quantum provides cloud-based quantum simulators such as ibmq$_{-}$qasm$_{-}$simulator, simulator$_{-}$statevector, and simulator$_{-}$mps \cite{balytskyi2023mathcal}. These simulators can be accessed by submitting quantum jobs from a local computer or an HPC-quantum hybrid system like the OLCF Frontier. We use these quantum simulators to conduct the performance analysis of QAOA, considering accuracy and time-to-solution.

We first run our QAOA jobs with various quantum simulators on the local computer. As can be seen in Figure 6A, all simulators quickly find accurate solutions for small-scale problems (sizes $\leq$ 10). Here, the depth of the ansatz circuit is 87 with the required number of qubits of 10. However, errors have increasingly occurred as the problem size grows. Notably, some local simulators (such as Aer/density matrix, statevector, and MPS) cannot handle large-scale problems (size $\geq$ 16; ansatz circuit depth $\geq$ 141, required qubits $\geq$ 16) due to their high memory requirements for simulating quantum behaviors \cite{pinto2022implementation}. BasicAer, Aer/automatic, ibmq$_{-}$qasm$_{-}$simulator, and simulator$_{-}$statevector simulators have better performance than other simulators in handling larger problems. Especially, ibmq$_{-}$qasm$_{-}$simulator and simulator$_{-}$statevector exhibit higher accuracies compared to the local simulators. The accuracy for a problem size of 20 (ansatz circuit depth: 177, required qubits: 20) is 0.99 (ibmq$_{-}$qasm$_{-}$simulator) and 0.99 (simulator$_{-}$statevector), which are higher than that of BasicAer (0.74) and Aer/automatic (0.79). Furthermore, ibmq$_{-}$qasm$_{-}$simulator and simulator$_{-}$statevector can be utilized for larger-scale problems, such as a size of 28, with high accuracy (0.98 and 0.99, respectively), which is beyond the limit of the local simulators. Although one local simulator (Aer/automatic) can solve large problems such as a problem size of 30, it exhibits a low accuracy (0.61).

Time-to-solution is also an important metric to evaluate the performance of quantum simulators for the quantum algorithm. Figure 6B indicates that the local simulators are much faster than the IBM simulators to handle small-scale problems (sizes $\leq$ 12). For example, the Aer/statevector achieves a time-to-solution of 4.07 s while the simulator$_{-}$statevector requires 735.16 s for a problem size of 12 (ansatz circuit depth: 105, required qubits: 12). This may be because queue time in the IBM server and communication time between the local computer and IBM server are significant. However, the benefit of using the IBM quantum simulators becomes apparent when dealing with larger problems. Many local simulators (Aer/density matrix, statevector, and MPS) have challenges in solving a problem size of 24 (ansatz circuit depth: 213, required qubits: 24), failing to solve the problem, and BasicAer and Aer/automatic require 7,725 s and 167 s, respectively. On the other hand, ibmq$_{-}$qasm$_{-}$simulator and simulator$_{-}$statevector can handle those problems within 820 s and 750 s, respectively. Given the lower accuracies of the local simulators (Aer/automatic), the IBM quantum simulators demonstrate much better performance. These results verify that the local quantum simulators are good for small problems (sizes $\leq$ 12), but the IBM quantum simulators have much better performance for larger problems. Nevertheless, it is observed that the time-to-solution for the IBM simulators intractably increases with increasing problem sizes, for example, the time-to-solution with simulator$_{-}$statevector increases from 750 s to 10,833 s when increasing problem size from 24 to 28 (ansatz circuit depth: 249, required qubits: 28). 

\begin{figure}[!ht]
\centering
\includegraphics[width=1.0\linewidth]{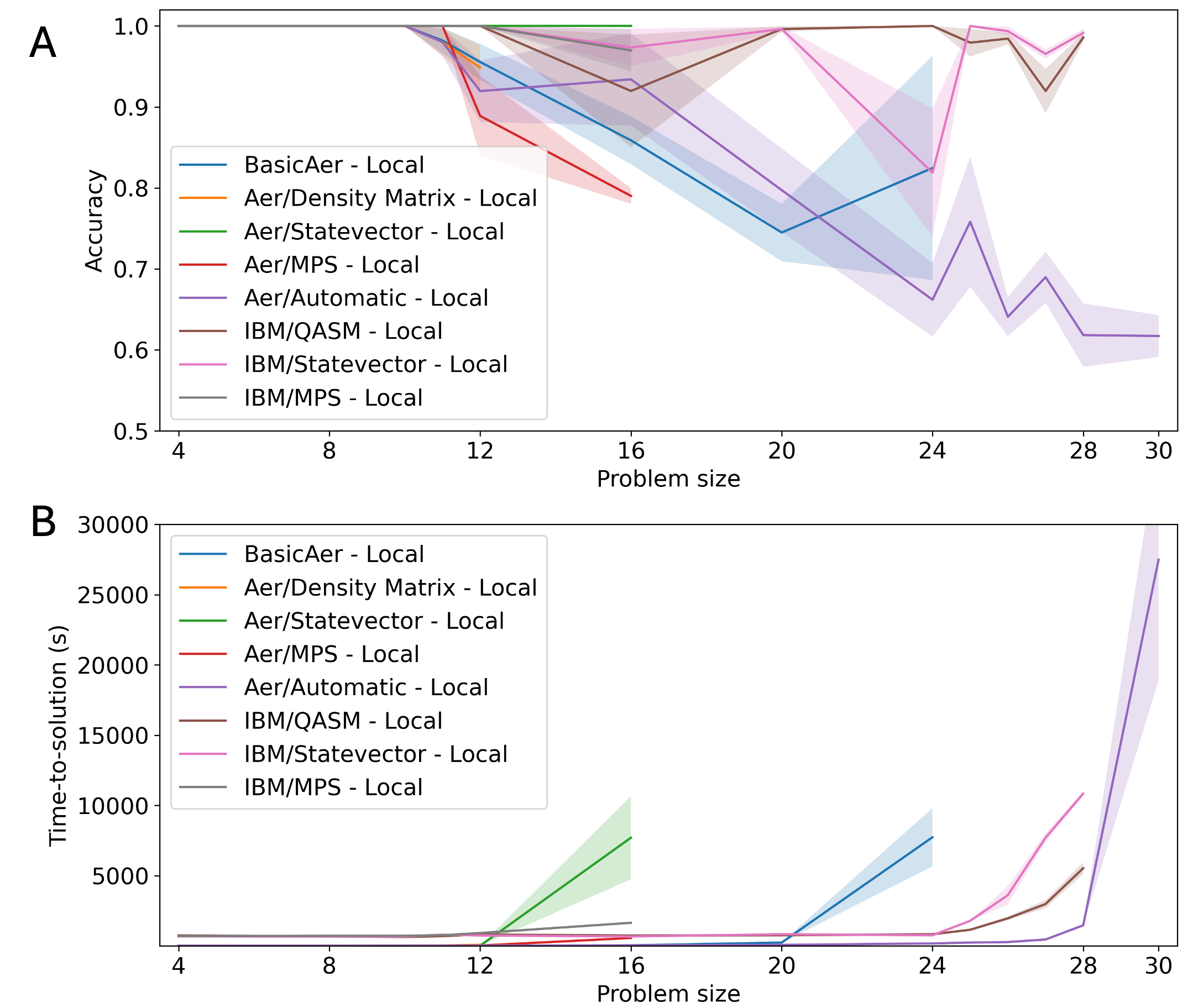}
\caption{\label{fig:fig-6}{Performance analysis of QAOA with different quantum simulators executed or submitted on the local computer. (A) Accuracy and (B) time-to-solution as a function of the problem size.}}
\end{figure}

We have shown the performance of the quantum simulators executed or submitted on the local computer. The performance can be enhanced when employing the HPC-quantum hybrid system (OLCF Frontier). To analyze the enhanced performance, we solve the same problems using QAOA with the same simulators, executed or submitted on the OLCF Frontier instead of on the local computer. The Frontier shows better capability to handle large problems, enabling the local simulators (Aer/density matrix and Aer/automatic) to solve problems with sizes of 16 and 32 respectively, which cannot be handled on the local computer. Despite its better capability, the Frontier cannot show a significant advantage in accuracy over the local computer because the mechanisms of the quantum simulators are basically the same (Figure 7A). However, the Frontier can reduce the time-to-solution, as can be seen in Figure 7B. The time-to-solution measured with the BasicAer, Aer/automatic, ibmq$_{-}$qasm$_{-}$simulator and simulator$_{-}$statevector to solve a problem size of 24 is 6,081 s, 118 s, 692 s, and 655 s, respectively. Furthermore, the time-to-solve measured with the simulator$_{-}$statevector to solve a larger problem (size: 28) can be greatly reduced from 10833 s to 4240 s, demonstrating approximately 155\% acceleration by submitting the quantum job from the HPC-quantum hybrid system. Moreover, considering the same type of quantum simulator (e.g., statevector), solving a problem (size: 16) takes 7,691 s on the local computer with Aer/statevector simulator. This time-to-solve greatly reduces to 702 s by solving the same problem on the Frontier with the same simulator (Aer/statevector). It can be further reduced to 314 s by submitting this job to simulator$_{-}$statevector on the HPC-quantum system, marking $\sim$24 times speedup compared to the local computer.

\begin{figure}[!ht]
\centering
\includegraphics[width=1.0\linewidth]{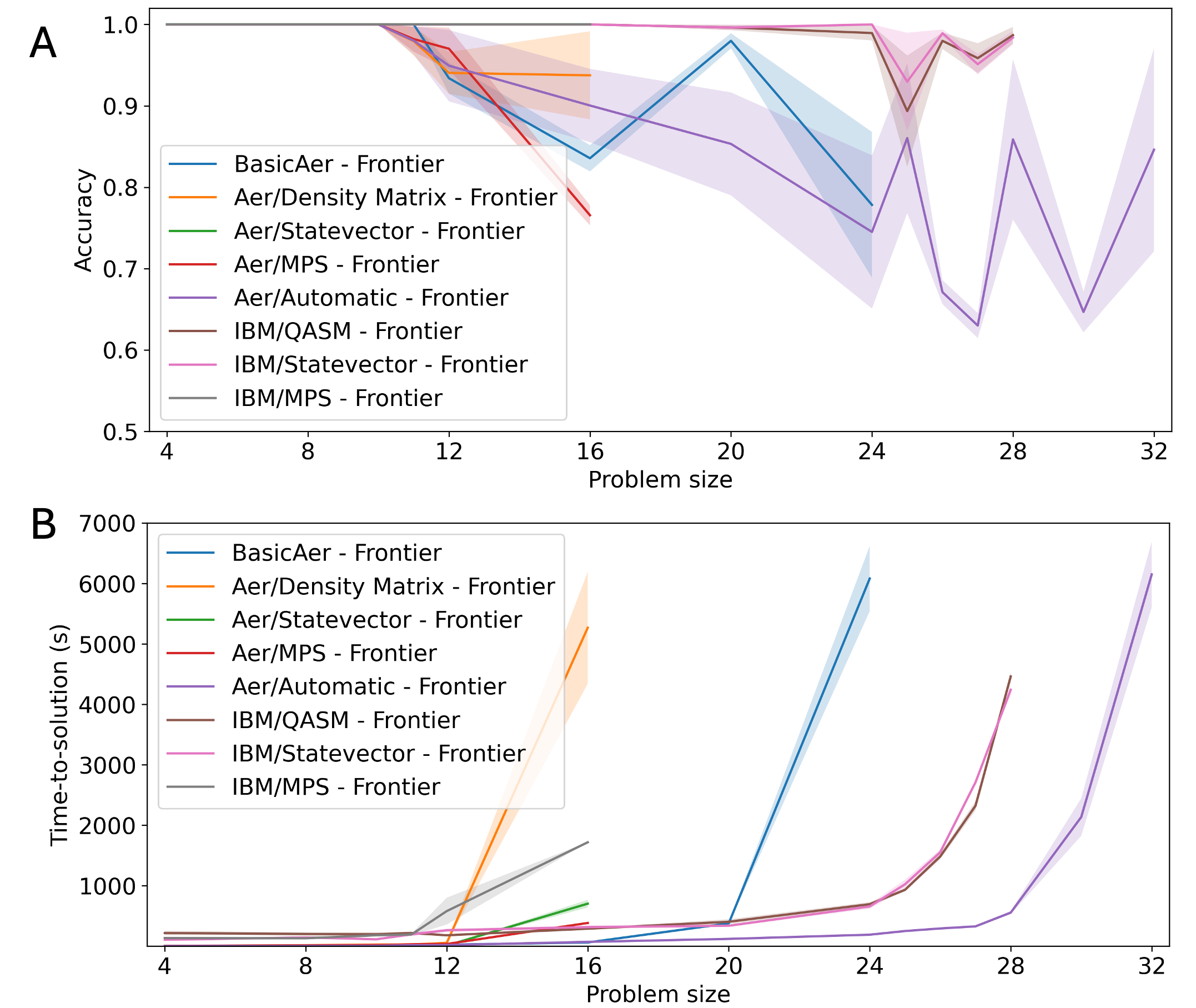}
\caption{\label{fig:fig-7}{Performance analysis of QAOA with different quantum simulators executed or submitted on an HPC-quantum hybrid system (Frontier). (A) Accuracy and (B) time-to-solution as a function of the problem size.}}
\end{figure}

These results clearly illustrate that the local quantum simulators can be better for solving small-scale optimization problems (sizes $\leq$ 12; ansatz circuit depth $\leq$ 105, required qubits $\leq$ 12) since they do not require any queue and communication times. However, for large-scale problems (sizes $\geq$ 16; ansatz circuit depth $\geq$ 141, required qubits $\geq$ 16), it has been demonstrated that QAOA jobs submitted on the HPC-quantum hybrid system to run IBM quantum simulators provide superior performance, presenting $\sim$24 times acceleration compared to the local computer. However, the accuracy of QAOA submitted either on the local computer or HPC-quantum system is at a similar level since the operational mechanisms of quantum simulators are the same. This analysis helps determine the most suitable quantum simulator for optimizing metamaterial structures with QAOA depending on the size of optimization spaces. Moreover, the integration of the efficient quantum simulator into the active learning algorithm will provide an efficient optimization process for metamaterial designs.

\subsection{Performance Analysis of HPC-Enhanced TMM}
We investigate the performance of HPC-enhanced TMM for simulating complex structures (1,000-layered structures) with numerous simulation conditions. We first study to determine the optimal number of processors used for MPI on the HPC system (OLCF Frontier) across different numbers of simulation conditions. As presented in Figure 8A, the required time for simulations varies depending on the number of processors used, with a notable decrease observed as more processors are used. The simulation time almost converges when using more than 20 processors; thus, we fix the number of processors to 20 for the rest of the TMM-related studies unless otherwise noted. 

Figure 8B illustrates the simulation time with different numbers of simulation conditions, run on a local computer with a single processor and the Frontier using MPI with 20 processors. When a computational workload is relatively low, such as a simulation number of 50, the local computer can solve the given optics problem in 18.13 s. However, the computation time grows rapidly on the local computer (from 18.13 to 2619.87 s) when increasing complexity (simulation numbers from 50 to 350). Note that the local computer cannot handle a highly complex case, such as a simulation number of 400 or more, due to the computational limitation. On the other hand, such complex problems can be efficiently solved on the Frontier owing to the parallel computing capability. It needs much less time than the local computer especially when involving numerous simulations; Simulation time linearly increases from 5.52 s to 39.16 s for the number of simulations increases from 50 to 350. The discrepancy in simulation time between the local computer and HPC system is exponentially increasing for highly complex problems, showing the HPC system’s remarkable acceleration capability from 328\% (50 simulation conditions) to $\sim$67 times (350 simulation conditions) compared to the local computer. 

Furthermore, the HPC system can solve problems beyond the limits that the local computer cannot handle, such as the number of simulation conditions of 400, 700, and 1,000. Figure 8C further demonstrates the HPC system’s linearly increasing time scale in the domain of large-scale problems. Assuming the active learning for the 1,000-layered structure requires 5,000 iterations where each iteration involves 10 s for ML and 1 s for solving a given surrogate model, it is estimated that total optimization takes 40.46 h on the local computer and 22.95 h on the Frontier (number of simulation conditions: 50). However, when considering 350 simulation conditions, the local computer requires 3,653.99 h, which is unrealistic, while the Frontier completes the take only in 69.67 h making $\sim$52 times speedup. Moreover, as shown in Figure 8D, the Frontier can be used for significantly large problems (e.g., number of simulation conditions: 400, 700, and 1,000), enabling the optimization to be completed in 174.91 h for the 1,000 case. 

\begin{figure}[!ht]
\centering
\includegraphics[width=1.0\linewidth]{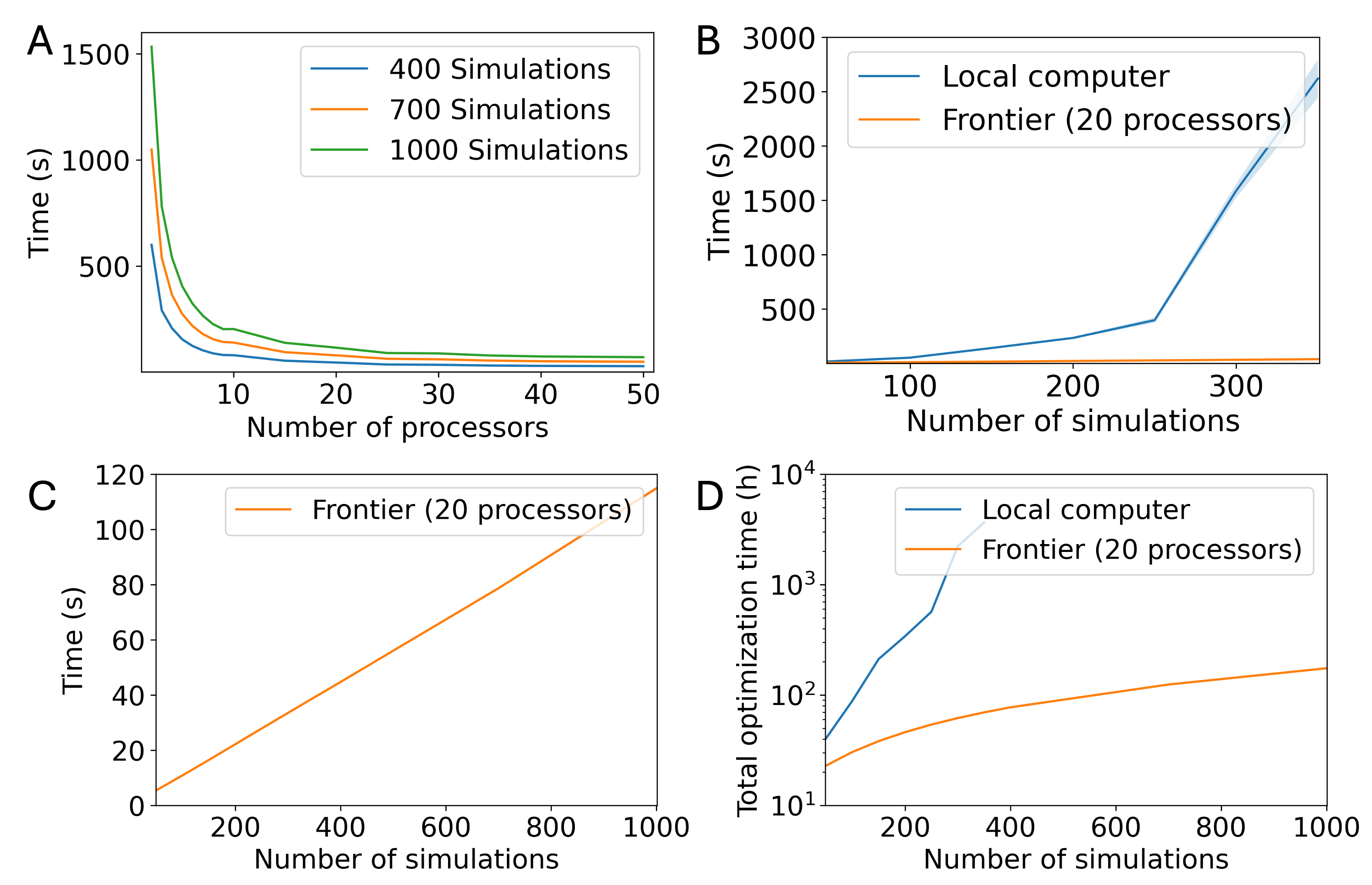}
\caption{\label{fig:fig-8}{Elapsed time to complete simulations. (A) Simulation time with different numbers of simulation conditions as a function of the number of processors, run on the Frontier. (B) Simulation time as a function of the number of simulation conditions, run on the local computer and Frontier (using MPI with 20 processors). (C) Simulation time on the Frontier for large-scale problems. (D) Estimated time to complete the iterative optimization process on the local computer and Frontier as a function of simulation numbers.}}
\end{figure}

We further test the performance of HPC-enhanced TMM using more nodes while keeping the total number of MPI processors to 20. Figure 9 indicates that HPC-enhanced TMM shows further improvement in simulation time under numerous simulation conditions by employing multi-nodes, showing $\sim$29\% acceleration. However, this improvement is not that dramatic compared to utilizing more processors, as we observed similar results in multi-node FM studies (Figure 5). Hence, it can be concluded that utilizing more MPI processors is the first option to accelerate TMM simulation on the HPC system, and using more nodes is the second option to further accelerate the simulation.

\begin{figure}[!ht]
\centering
\includegraphics[width=1.0\linewidth]{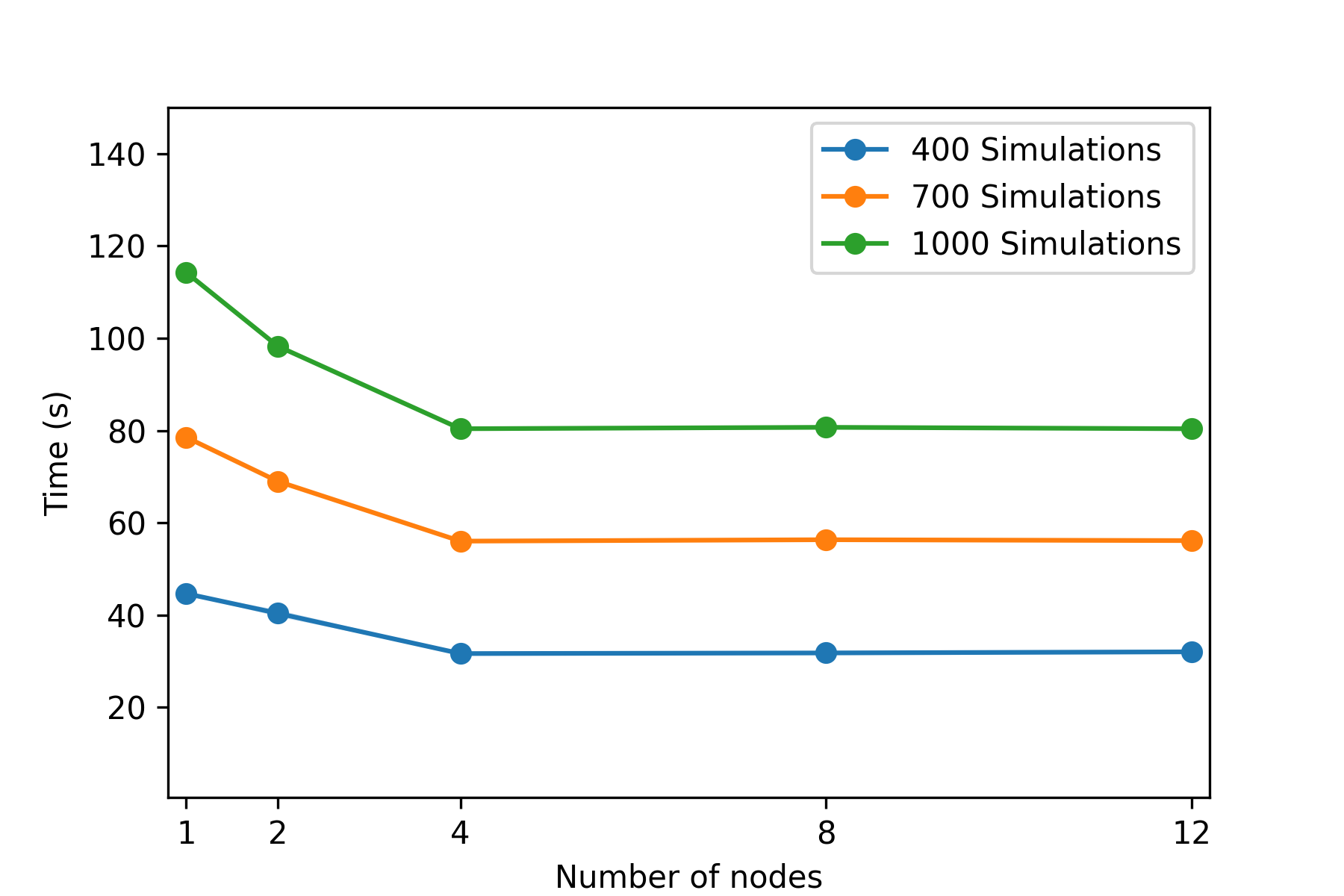}
\caption{\label{fig:fig-9}{Elapsed time to complete simulations when using more nodes while keeping the total number of MPI processors to 20.}}
\end{figure}

The results clearly illustrate the significant acceleration of wave-optics simulation ($\sim$67 times) achieved by the HPC-enhanced TMM using MPI parallelization. This acceleration not only enhances the efficiency of solving complex optimization problems but also enables the efficient handling of highly large-scale problems that are intractable in a conventional computing system. Therefore, researchers can tackle challenging optical simulation tasks with unprecedented speed and scalability by leveraging the computational power and parallel processing capabilities of HPC systems. As such, the integration of the HPC-enhanced simulation approach into the active learning algorithm will be crucial in overcoming the current limitations of optimization for highly complex metamaterials.  

\section{\textbf{Conclusion}}
In this work, we have analyzed the performance of the active learning algorithm for metamaterial design on the integrated HPC and quantum system. We have demonstrated the significant advantages in computational time by leveraging MPI parallelization on the HPC system, achieving $\sim$54\% acceleration for ML tasks and $\sim$67 times acceleration for wave-optics simulation. In addition, we have analyzed the performance of QAOA designed for solving combinatorial optimization problems with various quantum simulators. It should be noted that running the quantum algorithm on the integrated HPC and quantum hybrid system leads to $\sim$24 times speedup in solving a given combinatorial optimization problem. These findings highlight the efficacy of HPC-enhanced ML and simulation by taking advantage of MPI parallelization for distributing a heavy workload to multi-processors. Moreover, our results reveal the time advantages provided by the HPC-quantum system in solving optimization problems. Therefore, we expect that this optimization algorithm on the integrated HPC and quantum system will be applicable to various material designs that have highly complex and large design spaces, such as metamaterials, electronic devices, and energy materials.

\section{\textbf{Acknowledgment}}
This research used resources of the Oak Ridge Leadership Computing Facility at the Oak Ridge National Laboratory, which is supported by the Office of Science of the U.S. Department of Energy under Contract No. DE-AC05-00OR22725.
{\it Notice}: This manuscript has in part been authored by UT-Battelle, LLC under Contract No. DE-AC05-00OR22725 with the U.S. Department of Energy. The United States Government retains and the publisher, by accepting the article for publication, acknowledges that the U.S. Government retains a non-exclusive, paid up, irrevocable, world-wide license to publish or reproduce the published form of the manuscript, or allow others to do so, for U.S. Government purposes. The Department of Energy will provide public access to these results of federally sponsored research in accordance with the DOE Public Access Plan.


\bibliography{refs}
\bibliographystyle{IEEEtran}

\end{document}